\documentclass[10pt,aps,prx,twocolumn,superscriptaddress,floatfix,nofootinbib,natbib,longbibliography]{revtex4-2}
\usepackage{graphicx}
\usepackage{amsmath,amssymb,amsfonts,amsthm}
\usepackage{braket}
\usepackage[colorlinks=true,linkcolor=blue,urlcolor=blue,citecolor=blue,pdfusetitle]{hyperref}
\usepackage[dvipsnames]{xcolor}
\usepackage{dsfont}
\usepackage{booktabs}

\graphicspath{{figures/}}

\renewcommand{\sp}{\ensuremath{\sigma^+}}
\newcommand{\sm}{\ensuremath{\sigma^-}}

\renewcommand\vec{\boldsymbol}

\newcommand{\paulistring}[1]{\emph{#1}}
\newcommand{\kg}{\ensuremath{\kappa_0 / g - \kappa_1 / g}}
\newcommand{\param}{\ensuremath{K}}

\begin{document}
\title{Variational Quantum Simulation of the Interacting Schwinger Model on a Trapped-Ion Quantum Processor}
\author{Christian Melzer}\affiliation{Quantum, Institut for Physics, Johannes Gutenberg-Universität Mainz}
\author{Stephan Schuster}\affiliation{Quantum Optics and Quantum Information Group, Friedrich-Alexander-Universität Erlangen-Nürnberg}
\author{Diego Alberto Olvera Millán}\affiliation{Quantum, Institut for Physics, Johannes Gutenberg-Universität Mainz}
\author{Janine Hilder}\affiliation{Quantum, Institut for Physics, Johannes Gutenberg-Universität Mainz}
\author{Ulrich Poschinger}\affiliation{Quantum, Institut for Physics, Johannes Gutenberg-Universität Mainz}
\author{Karl Jansen}\affiliation{Computation-Based Science and Technology Research Center, The Cyprus Institute, 20 Kavafi Street, 2121 Nicosia, Cyprus}
\affiliation{Deutsches Elektronen-Synchrotron DESY, Platanenallee 6, 15738 Zeuthen, Germany}
\author{Ferdinand Schmidt-Kaler}\affiliation{Quantum, Institut for Physics, Johannes Gutenberg-Universität Mainz}

\date{\today}
\begin{abstract}
Simulations in high-energy physics are currently emerging as an application of noisy intermediate-scale quantum (NISQ) computers. In this work, we explore the multi-flavor lattice Schwinger model -- a toy model inspired by quantum chromodynamics -- in one spatial dimension and with nonzero chemical potential by means of variational quantum simulation on a shuttling-based trapped-ion quantum processor. This fermionic problem becomes intractable for classical numerical methods even for small system sizes due to the notorious sign problem. We employ a parametric quantum circuit executed on our quantum processor to identify ground states in different parameter regimes of the model, mapping out a quantum phase transition which is the hallmark feature of the model. The resulting states are analyzed via quantum state tomography, to reveal how characteristic properties such as correlations in the output state change across the phase transition. Moreover, we use the results to determine the phase boundaries of the model.
\end{abstract}
\maketitle

\section{Introduction}

Variational methods of quantum computing are of increasing interest for the investigation of lattice field theories in order to find new avenues for numerical simulations addressing models and regimes where classical Monte Carlo techniques break down~\cite{Banuls2020,Bauer:2022hpo,Funcke2023a,DiMeglio2023}. While current noisy intermediate-scale quantum (NISQ) devices still suffer from hardware limitations such that classical numerical simulations still outperform them~\cite{Banuls2018a,Banuls2019,Schuster_2024}, the ongoing rapid maturation of quantum hardware will eventually allow to quantitatively explore sign afflicted problems in  lattice field theory. The last years have seen an increasing number of proof-of-principle demonstrations for simulations of non-trivial problems from high-energy physics on actual quantum computer hardware~\cite{barends2015digital,martinez2016real,kokail2019self,chertkov2021holographic,Meth2022,mueller2023quantum,Meth2025}.

Common methods for the investigation of lattice field theories on quantum computers are based on the variational quantum eigensolver (VQE) approach. The VQE protocol tries to find the approximate ground state of a given Hamiltonian using a parameterized quantum circuit in a closed feedback loop with a classical optimizer. The cost function is most often the measured energy expectation value of the quantum state prepared by the ansatz circuit for a given fixed parameter set. The parameters of the quantum circuit are then optimized until the specified convergence criteria are met. VQE, in combination with circuit optimization and error mitigation techniques, has been shown to be well-suited for NISQ devices~\cite{Lorenza1998, Endo2018, Funcke2021a, Cai2020, Giurgica-Tiron2020, Berg2023, Peruzzo2014, McClean2016}. 

In this work, we implement the parameterized quantum circuit originally conceived in~\cite{Schuster_2024} on a 4-qubit trapped-ion quantum processor. We perform full VQE runs with this circuit, determining the ground-state energy of the two-flavor Schwinger model in the presence of a chemical potential. This model represents an interesting toy model from lattice gauge theory which shares many similarities to quantum chromodynamics~\cite{Lohmayer2013, Narayanan2012, Banuls2016a, Coleman:1976uz, Funcke2019}. The model is known to exhibit an infinite number of first-order energy phase transitions as one varies the chemical potentials. However, being restricted to 4 qubits, the exploration within this work is restricted to three phases of the system. In particular, we show that our VQE algorithm converges well in three neighboring phases, allowing us to also determine the phase boundaries. Additionally, we perform a full quantum state tomography for one parameter set within each phase. 

The paper is organized as follows: In Sec.~\ref{sec:target_model}, we present the target model for our VQE and its relevant properties, summarizing the more detailed discussion of the model in~\cite{Schuster_2024}. We continue by presenting the relevant details of our trapped-ion quantum processor in Sec.~\ref{sec:hardware_system}. Next, we describe the quantum circuit which we use for the VQE algorithm in Sec.~\ref{sec:procedures_methods}, along with details on its execution environment, i.\,e. the required measurements, the execution parameters, and the classical optimizer. The following Sec.~\ref{sec:results} presents the experimental results, before we summarize our findings in Sec.~\ref{sec:assessment}.

\section{The Multi-Flavor Schwinger Model}
\label{sec:target_model}

\begin{figure}[h]
    \centering
    \includegraphics[width=\linewidth]{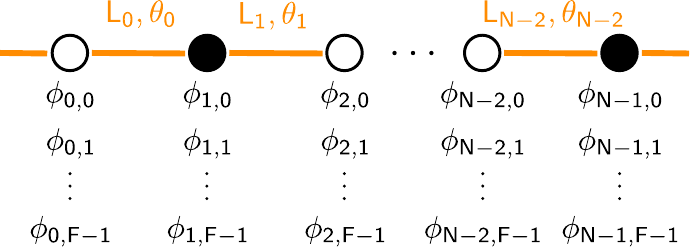}
    \\[1em]
    \includegraphics[width=0.6\linewidth]{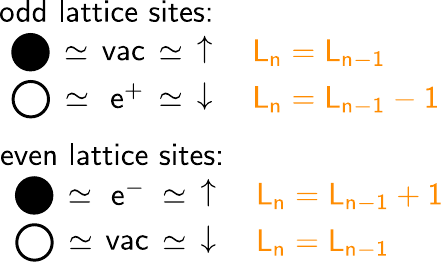}
    \caption{Lattice encoding of the multi-flavor Schwinger model. 
    The single-component fermionic fields $\phi_{n,f}$ are defined on the staggered lattice sites $n$ (filled and unfilled circles, left to right). At each lattice site, $F$ single-component fields are defined with different flavors $f$ (top to bottom). At neighboring lattice sites $n$ and $n+1$, the fields couple via the operators $L_n$ and $\theta_n$. According to Gauss law (cf. Eq.~\eqref{eq:gauss_law}), occupied even lattice sites translate to the presence of an particle and unoccupied odd lattice sites to the presence of an anti-particle, which is here shown in an example configuration.}
    \label{fig:mfs_lattice_mapping}
\end{figure}

We investigate the multi-flavor Schwinger model with staggered fermions in its Hamiltonian lattice formulation~\cite{Banuls2016a, Schuster_2024},
\begin{align}
    \begin{aligned}
         H = &-\frac{i}{2a}\sum_{n=0}^{N-2}\sum_{f=0}^{F-1}\left(\phi^\dagger_{n,f}e^{i\theta_n}\phi_{n+1,f}-\mathrm{h.c.}\right)\\
         &+\sum_{n=0}^{N-1}\sum_{f=0}^{F-1}\left(m_f(-1)^n +\kappa_f \right)\phi^\dagger_{n,f}\phi_{n,f}\\
         &+ \frac{g^2a}{2}\sum_{n=0}^{N-2} L_n^2,
    \end{aligned}
    \label{eq:hamiltonian}
\end{align}
with $F$ fermion flavors and $N$ lattice sites of spacing $a$. Within this formulation, we use the Kogut-Susskind lattice mapping, which splits the Dirac bispinor and maps the particle and anti-particle fields of all flavors onto  even and odd lattice sites, respectively~\cite{Kogut1975}. The single-component fermionic field of flavor $f$ on site $n$ is denoted $\phi_{n,f}$, satisfying the usual canonical anti-commutation relation $\{\phi^\dagger_{n,f},\phi^{\phantom{\dagger}}_{n',f'}\}=\delta_{n,n'}\delta_{f,f'}$. The electric field operator $L_n$ and its canonical conjugate $\theta_n$ act on the links between the staggered lattice sites $n$ and $n+1$ (cf. Fig.~\ref{fig:mfs_lattice_mapping}). The coupling parameter is denoted by $g$ and $m_f$, $\kappa_f$ correspond to the bare mass and the bare chemical potential of flavor $f$, respectively. All physical states of the above Hamiltonian have to follow Gauss' law in its staggered lattice version:
\begin{equation}
    L_n-L_{n-1} = \sum_{f=0}^{F-1}\phi^\dagger_{n,f}\phi_{n,f}-\frac{F}{2}\left(1-(-1)^n\right).
    \label{eq:gauss_law}
\end{equation}
Excitations of a flavor $f$ on even (odd) lattice sites thus create $+1$ ($-1$) units of flux, which correspond to particle (anti-particle) excitations (cf. Fig.~\ref{fig:mfs_lattice_mapping})~\cite{Hamer1997}.
The right-hand side of Eq.~\eqref{eq:gauss_law} denotes the staggered charge $Q_n$~\cite{Schuster_2024},
\begin{equation}
    Q_n = \sum_{f=0}^{F-1}\phi^\dagger_{n,f}\phi_{n,f}-\frac{F}{2}\left(1-(-1)^n\right).
    \label{eq:stagCharge}
\end{equation}
For our VQE, we perform a residual gauge transformation to eliminate the gauge field $e^{i\theta_n}$ and apply a Jordan-Wigner transformation, mapping the fermionic degrees of freedom onto spins~\cite{Schuster_2024}. Concretely, we map the staggered lattice sites $n$ and fermion flavors $f$ onto qubits $p$ according to 
\begin{equation}
    p=nF+f.
    \label{eq:JWqubitmapping}
\end{equation}
Additional rescaling yields the dimensionless Hamiltonian
\begin{align}
    \begin{aligned}
         W =& -ix\sum_{n=0}^{N-2}\sum_{f=0}^{F-1}\bigg(\sp_{n,f}(iZ_{n,f+1})\dots(iZ_{n+1,f-1}) \sm_{n+1,f}\\
         &-\sm_{n,f}(-iZ_{n,f+1})\dots(-iZ_{n+1,f-1}) \sp_{n+1,f}\bigg)\\
         &+\sum_{n=0}^{N-1}\sum_{f=0}^{F-1}\left(\mu_f(-1)^n +\nu_f \right)\frac{1}{2}\left(Z_{n,f} + \mathds{1}\right) \\
         &+ \sum_{n=0}^{N-2} \left( \sum_{k=0}^n Q_k\right)^2,
    \end{aligned}
    \label{eq:hamiltonian_dimensionless}
\end{align}
where $X$, $Y$ and $Z$ are the usual Pauli matrices and $\sigma^\pm \equiv (X \pm i Y)/2$ are jump operators. The Hamiltonian is now parameterized by the dimensionless quantities $x \equiv 1/(ag)^2$, $\mu_f \equiv 2\sqrt{x}m_f/g$, $\nu_f \equiv 2\sqrt{x}\kappa_f/g$. In this work, we consider a charge neutral situation and thus require that $Q_\mathrm{tot}\ket{\psi} = 0$ with $Q_\mathrm{tot} = \sum_{n=0}^{N-1}Q_n$ and $Q_n$ from Eq.~\eqref{eq:stagCharge}. Additionally, we focus on the minimal non-trivial example of two fermion flavors $F=2$ and two lattice sites $N=2$. The four required qubits map to the lattice indices and flavor indices according to Eq.~\eqref{eq:JWqubitmapping}. For a derivation of the form of Eq.~\eqref{eq:hamiltonian_dimensionless} and more details about the model and the following theoretical results, we refer to \cite{Schuster_2024}.

The model exhibits an infinite number of first-order phase transitions, if we change the chemical potential of one flavor $f$ while fixing the chemical potential of another flavor $f'$. The phase transitions are thus controlled by the chemical potential difference $\nu_f-\nu_{f'}$, and the different phases are characterized by the number of particles of flavor~$f$:
\begin{equation}
    \label{eq:particle-number-by-flavor}
    N_f = \sum_{n=0}^{N-1}\frac{1}{2}\left(Z_{n,f}+\mathds{1}\right),
\end{equation}
Each phase is therefore labeled by the particle number vector $\boldsymbol{N}=\left(N_0,N_1\right)^T$. Since all particle numbers $N_f$ commute with the Hamiltonian $W$, we can write the ground state energy of each phase in the following form
\begin{equation}
    \label{eq:energyPT}
    E_{\boldsymbol{N}}(\boldsymbol{\nu}) =\boldsymbol{\nu}\cdot\boldsymbol{N} + E^{\text{min}}_{\boldsymbol{N}},
\end{equation}
where the effective chemical potentials were also bundled into a vector $\boldsymbol{\nu}$ and $E^{\text{min}}_{\boldsymbol{N}}$ is an energy offset which is constant across each phase. By measuring $E_{\boldsymbol{N}}(\boldsymbol{\nu})$ as well as all particle numbers $N_f$ at one particular point $(\boldsymbol{\nu})$, we can calculate this constant for the whole phase. Right at the critical point of a first-order energy phase transition, the ground state energy of the two neighboring phases $\boldsymbol{N}$ and $\bar{\boldsymbol{N}}$ is degenerate, $E_{\boldsymbol{N}}(\boldsymbol{\nu})=E_{\bar{\boldsymbol{N}}}(\boldsymbol{\nu})$. From this equality, Eq.~\eqref{eq:energyPT} and $N=N_0+N_1$, we derive the following expression for the critical point for the case of two flavors $F=2$:
\begin{equation}
    \label{eq:MFS_crit_pt}
    (\nu_0-\nu_1)\lvert_{\mathrm{crit}} = \frac{E_{\boldsymbol{N}'}(\boldsymbol{\nu}')+\boldsymbol{\nu}'\cdot\boldsymbol{N}'-E_{\boldsymbol{N}''}(\boldsymbol{\nu}'')+\boldsymbol{\nu}''\cdot\boldsymbol{N}''}{N_0''-N_0'}
\end{equation}
The energies $E_{\boldsymbol{N}}(\boldsymbol{\nu}')$, $E_{\boldsymbol{\bar{N}}}(\boldsymbol{\nu}'')$ are measured for two different chemical potential vectors $\boldsymbol{\nu}'$, $\boldsymbol{\nu}''$ which lie within two different but neighboring phases, and $\boldsymbol{N}'$, $\boldsymbol{N}''$ are the measured particle numbers for these phases.
The expression in Eq.~\eqref{eq:MFS_crit_pt} will be used later on to calculate the critical points between the investigated phases from our VQE results.\\
In the following, we present details of the platform used to run VQE simulation of the model, in terms of its hardware and software environments.

\begin{figure*}
    \includegraphics[width=\linewidth]{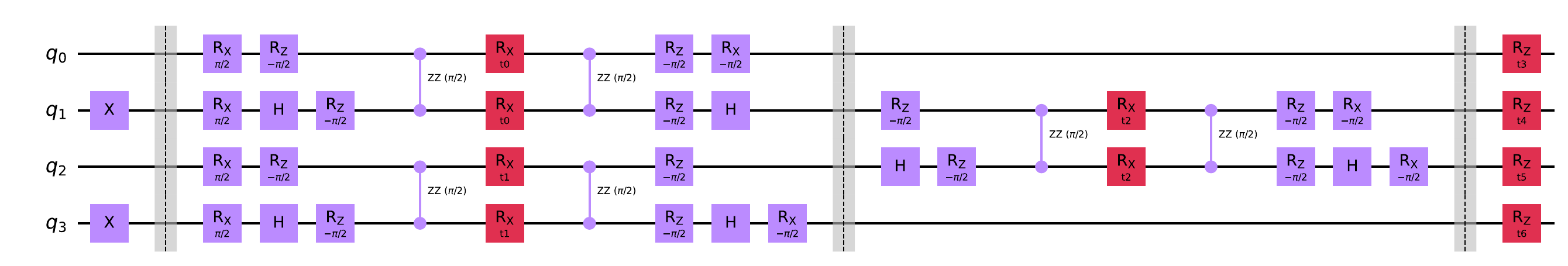}
    \caption{Parameterized circuit for variational exploration of the Hamiltonian Eq.~\eqref{eq:hamiltonian_dimensionless}. Boxes indicate various single qubit gate operations, in particular parameterized gates (red) and fixed gates (magenta). Two-qubit entangling gate operations are implemented as Ising-type gates $ZZ(\pi/2)$. The two Pauli $X$ gates before the first barrier initialize a charge-neutral state. The gates between the first and second barriers realize the parametric exchange gates (cf. Eq.~\eqref{eq:xx-yy-gate}) $U^{xy}_{01}(\theta_0)$ and $U^{xy}_{23}(\theta_2)$, and the gates between the second and third barriers realize $U^{xy}_{12}(\theta_1)$. Note the some local gates on qubits 1 and 2 cancel out around the second barrier. The gates beyond the third barrier are the $z$ rotations Eq.~\eqref{eq:z-rotation_def}. The local basis rotations and subsequent measurements are not shown. }
    \label{fig:heqs-circuit}
\end{figure*}

\section{Trapped-Ion Quantum Processor}
\label{sec:hardware_system}

For executing the digital quantum simulation of the two-flavor Schwinger model, we employ a quantum processor based on trapped-ion qubits stored in a segmented radiofrequency trap. We follow the shuttling-based approach \cite{Kielpinski2002,kaushal2020,QuantinuumRaceTrack} where small numbers of ion qubits are simultaneously stored  at one trap segment, in the form of linear Coulomb crystals. Several such crystal can be stored at different trap sites, and the storage configuration can be dynamically changed throughout the runtime of a quantum circuit. These shuttling operations are realized by applying suitable voltage waveforms to the trap electrodes, and consist of transport of ions between segments, ion crystal rotation of storage order reversal, and splitting and merging of ion crystals. Laser driven state preparation and readout (SPAM) as well as gate operations are carried out at a fixed trap site termed laser interaction zone (LIZ).  Throughout the execution of a circuit, qubits which are not involved in a gate operation are stored in other locations of the segmented trap. Execution of a circuit thus consists of alternating gate operations in the LIZ with shuttling operations for storage reconfiguration.

Our qubit is encoded in the two Zeeman sublevels of the S$_{1/2}$ ground state of $^{40}$Ca$^+$ ions, split by an external bias magnetic field of 0.37\,mT or respectively by about $2\pi\times$10\,MHz. We observe a spin-echo coherence time of up to 2.1\,s~\cite{Ruster2016LongLived}, limited by ambient magnetic field fluctuations. All gate operations are driven by laser beams, detuned from the $S_{1/2}\leftrightarrow P_{1/2}$ electric dipole transition near 397\,nm by about 800\,GHz. Local qubit rotations are realized via stimulated Raman transitions, driven by a pair of co-propagating beams, allowing for rotations of the form $R(\theta,\phi)=\exp\left[-\tfrac{i}{2}\theta\left(\cos\phi\;X+\sin\phi\;Y\right)\right]$ with freely variable angles $\theta,\phi$. Local rotations about $Z$, i.\,e. $R_z(\theta)=\exp\left[-\tfrac{i}{2}\theta\;Z\right]$ are carried out in a purely virtual manner: The cumulated rotation angles are tracked for each qubit, and the phase angles $\phi$ of subsequent rotations are corrected by these cumulated angles. Parasitic $Z$-rotations incur from shuttling, because the quantizing magnetic field is inhomogeneous across the trap structure. The resulting phases are also tracked and compensated for in the same way as the virtual gates. For a single qubit, randomized benchmarking reveals single-qubit gate errors as low as 7(2)$\times 10^{-4}$. Note that simultaneous local rotations can be carried out on two qubits stored in the LIZ. The native two-qubit entangling gate operation of our platform is a Ising-type phase gate between qubit $i,j$ of the form $G_{ij}=\exp\left[-i\tfrac{\pi}{4} Z_{i} Z_{j}\right] X_{i} X_{j}$, mediated by spin-dependent optical dipole forces \cite{Hilder2022FTR} acting on transverse oscillation modes of a two-ion crystal. The additional local $X$ operators arise from rephasing pulses used within the gate to mitigate errors and improve overall qubit coherence.  While this gate requires laser cooling close to the motional ground states of the transverse modes of the participating qubits, it is insensitive to motion along the axial direction, which is the direction mostly affected by the shuttling operations. Thus, parasitic motional excitation incurred from shuttling does not affect the fidelity of subsequent entangling gates for sufficiently small circuits, such that sympathetic cooling throughout the circuit runtime is not required. The fidelity of the two-qubit gate, estimated via monitoring the decay parity oscillation contrast versus number of subsequent gates, is about 99.0(2)\,\%, mainly limited by spontaneous photon scattering from the gate laser beams.\\
Qubit readout is carried out by shelving of population from on qubit state to the metastable $D_{5/2}$ state via driving a narrow electric quadrupole transition, followed by detection of state-dependent fluorescence. We observe combined SPAM errors of about 0.5(2)\,\%, mainly limited by shuttling-induced motional excitation affecting the electron shelving.

Trapped-ion platforms achieve relatively slow operation speeds, in our case leading to about 40\,ms per shot required for executing the VQE circuit shown in Fig. \ref{fig:heqs-circuit} and discussed further below. Especially in the context of VQE, this leads to rather long total data acquisition times, over which a sufficiently accurate calibration of the system needs to be maintained by a combination of passive and active measures. The latter consists of employing an automatic calibration routine for all calibrations that are to be repeatedly preformed throughout the data acquisition. This includes recalibration of atomic resonances and pulse areas (in terms of pulse duration or power), as well as shim voltages used for precise ion positioning. The latter most notably includes an axial electric bias field for splitting and merging of ion crystals and compensation of undesired micro-motion. 

The execution of the hybrid classical/quantum VQE algorithm on the quantum backend is enabled by a layered software environment. It allows the submission of circuits in \texttt{OpenQASM}~3 format \cite{OpenQASM3} and the retrieval of the result through HTTPS communication by the classical client side. A circuit submitted for execution is internally checked for correctness and system compliance and then compiled in two stages: First, the circuit is transpiled to the native gate set of the quantum backend and further optimized on the gate-level. The developed circuit compiler employs \texttt{pytket} compilation passes and custom phase-tracking, as well as some commutation techniques for optimization \cite{Kreppel2023}. Subsequently, the circuit is mapped to our hardware system. As the shuttling operations incur a considerable timing overhead \cite{Hilder2022FTR}, an additional shuttling compiler \cite{Janis2022} generates an optimized sequence of shuttling operations -- the shuttling schedule -- for enabling the execution of the desired gate sequence at minimized shuttling overhead.  The resulting sequence of hardware operations defines the execution of the circuit via a real-time hardware control systems \cite{kaushal2020}. The results are collected and stored in a database.

\section{Procedures and Methods}
\label{sec:procedures_methods}

\subsection{VQE execution}

The VQE protocol employs a parameterized ansatz quantum circuit in a closed feedback loop with a classical optimizer to approximate the ground state of the  model Hamiltonian $W$ in Eq.~\eqref{eq:hamiltonian_dimensionless}. The ansatz circuit, executed with a parameter vector $\boldsymbol{\theta}\in\mathbb{R}^p$, ideally produces a trial state $\ket{\Psi(\boldsymbol{\theta})}$ from a fixed initial state $\ket{\Psi_0}$. From a set of i.\,i.\,d. preparations of the trial state, we estimate the energy $\braket{\Psi(\boldsymbol{\theta})\lvert W\lvert \Psi(\boldsymbol{\theta})}$ via measurement of different observables. The classical optimizer evaluates this energy value as a cost function and tries to find the parameter vector $\boldsymbol{\theta}_{\mathrm{opt}}$ for which the energy value is a global minimum. The resulting state $\ket{\Psi(\boldsymbol{\theta}_{\mathrm{opt}})}$ is then the approximate ground state of $W$.\\
The parameterized ansatz quantum circuit employed in this work is shown in Eq.~\eqref{eq:vqe_circuit}. Its parametric is comprised of local $Z$-rotations $ R_i^z(\theta)$ and fermionic exchange gates $U_{ij}^{xy}(\theta)$ \cite{Anselmetti_2021} between qubits $i,j$, and leads to the following unitary evolution: 
\begin{equation}
    \label{eq:vqe_circuit}
    U(\boldsymbol{\theta})=\prod_{i=0}^{NF-1}R^z_i(\theta_{NF-1+i})\prod_{i \text{ odd}}^{NF-2}U^{xy}_{i,i+1}(\theta_i)\prod_{i \text{ even}}^{NF-2}U^{xy}_{i,i+1}(\theta_i),
\end{equation}
where
\begin{align}
    U_{ij}^{xy}(\theta)&=e^{-i\frac{\theta}{2}(X_iX_{j}+Y_iY_j)},\label{eq:xx-yy-gate}\\
    R_i^z(\theta)&= e^{-i\frac{\theta}{2}Z_i}. \label{eq:z-rotation_def}
\end{align}
For this work, we restricted our investigations to the minimal example of four qubits ($N=2$, $F=2$) and use only a single layer of the ansatz circuit that was originally presented in~\cite{Schuster_2024}. As further system parameters, $x = 16.0$, $m_0/g = m_1/g = 0$ and $\kappa_1 / g = 0 = \nu_1$ were chosen, while $\kappa_0 / g$ is varied.
The unitary operation of Eq.~\eqref{eq:vqe_circuit} is applied to the initial state $\ket{\Psi_0}$, which prepares the trial state
\begin{equation}
    \ket{\Psi(\vec{\theta})} = U(\vec{\theta})\ket{\Psi_0}
\end{equation}
The ansatz Eq.~\eqref{eq:vqe_circuit} is comprised only of gates which conserve the total charge $Q_{\mathrm{tot}}$ (cf. Sec.~\ref{sec:target_model}), as can be seen from Eq.~\eqref{eq:xx-yy-gate} and Eq.~\eqref{eq:z-rotation_def}, showing that the generators commute with the total charge operator. We thus ensure that the VQE search of the approximate ground state is restricted to the zero total charge subspace by choosing the initial state to have also vanishing total charge $\ket{\Psi_0}=\ket{0101}$. \\
As shown earlier in~\cite{Schuster_2024} via ideal simulations, this circuit can approximate the ground state of a similar, three-flavor version of our model in different parameter regimes. However, due to the high noise level on the considered IBM hardware, performing full VQE runs in the investigated parameter regimes were not possible. In this work, we show that full VQE runs with the described circuit for a two-flavor model also converge on an actual quantum hardware backend, given that comparatively small noise level of the device.

For the classical part of the VQE, we chose the SPSA optimization algorithm \cite{spall1992multivariate}, which performs well on noisy cost function evaluations. In every VQE run, we pick the initial angle parameters uniformly at random. The SPSA optimizer is initialized with three random parameter vectors for approximating the cost gradient, as well as an additional sampling for the energy value determination. For each trial parameter vector, the Pauli strings \paulistring{IXZY}, \paulistring{YZXI}, \paulistring{XZYI}, \paulistring{IYZX}, \paulistring{IIII}, \paulistring{ZIII}, \paulistring{IZII}, \paulistring{IIZI}, \paulistring{IIIZ} and \paulistring{ZZII} have to be measured, as obtained from writing out the model Hamiltonian Eq.~\eqref{eq:hamiltonian_dimensionless} in terms of Pauli operators. The Pauli strings are conveniently combined into the Pauli strings \paulistring{IXZY}, \paulistring{YZXZ}, \paulistring{XZYI}, \paulistring{ZYZX} and \paulistring{ZZII}, each of which is measured with 100 shots per evaluation.\\
A single run for determining viable SPSA hyperparameters with the Qiskit-inbuilt calibration functionality was executed on the quantum backend and used for all VQE runs \cite{footnote1}.

The ansatz used for this circuit is already mostly formulated in terms of native gates of the used quantum backend. The circuit compiler only decomposes the Hadamard gates and subsequently fuses local rotations to reduce the overall gate count.
As part of the compilation flow, a gate execution order is determined, which retains the desired unitary and leads to a favorable sequence of shuttling operations. The resulting shuttling schedule contains 33 gate operations, about 120 shuttles, and 16 split or merge operations.

Executing this unitary part of the circuit on the quantum processor requires about 10\,ms. Since the circuit runs were spread over different days and recalibrations, with recalibration measurements interleaved with the data acquisition, the effective shot rates slightly deviate. Including SPAM operations, the average time required per shot is about 40\,ms. The evaluation of one trial parameter vector with 100 shots per Pauli string thus adds up to 20 seconds of raw execution time. Additional overhead from software, retrapping upon ion loss and regular automatic recalibration routines increases the time per measurement point to 100 seconds. Retrapping lost ions usually requires less than ten seconds.

Using VQE, we explore the ground state of the $W$ for multiple values of 
\begin{equation}
\param = \kg
\end{equation}
and determine the energy for these points as well as calculating the phase transitions using Eq.~\eqref{eq:MFS_crit_pt} and an additional measurement of the particle number for each of the points. The three chosen parameter values are $\param = -14$, $\param = 0$ and $\param = 10$.

\subsection{State tomography}

For the final parameter set retrieved by the respective VQE runs for each value of $K$, we perform quantum state tomography \cite{QiskitStateTomography} to reconstruct the resulting state. The circuit from Fig.~\ref{fig:heqs-circuit} is used with the optimum parameter set, and all 81 four-qubit Pauli strings not containing at least one identity are measured using 400 shots each. Such a tomography measurement takes around two hours on the quantum backend and yields an estimated density matrix $\rho(\boldsymbol{\theta}_{\text{opt}})$ via a maximum-likelihood estimate using linear inversion \cite{QiskitLinearInversion}. For reference, the exact state $\ket{\Psi(\boldsymbol{\theta}_{\text{opt}})}$ pertaining to the optimum circuit parameters is calculated on a simulation backend. The overlap of the reconstructed density matrices with the reference state is calculated as $\mathcal{F}=\braket{\Psi(\boldsymbol{\theta}_{\text{opt}})|\rho(\boldsymbol{\theta}_{\text{opt}})|\Psi(\boldsymbol{\theta}_{\text{opt}})}$.\\
The different phases can be distinguished by their correlations between the staggered lattice sites and between the flavor indices, which can be quantified by the quantum mutual information of different two two-qubit partitions of the system. Uncertainties for all quantities computed from estimated density matrices are obtained via parametric bootstrapping: Artificial tomography data is numerically sampled based on the reconstructed $\rho$, and an artificial tomography result $\rho°$ is reconstructed from this data. The metric of interest is then calculated for many samples of $\rho°$ to gather statistics of any metric.

\section{Results}
\label{sec:results}

\begin{figure}
    \centering
    \includegraphics[width=1.0\linewidth]{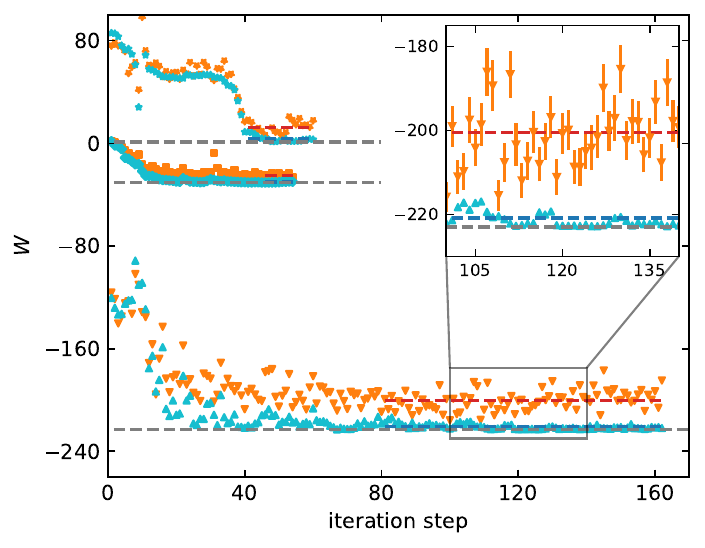}
    \caption{VQE runs for $\param$ are 10, 0 and -14, from top to bottom. Using a set of seven parameters for the parameterized gates (see Fig.~\ref{fig:heqs-circuit}), the quantum circuit is executed on the trapped ion quantum processor to determine the energy value. From averaging five circuit evaluations with 100 shots each we evaluate $W$ on the quantum hardware (from top to bottom: orange stars with point down, squares and triangles with point down). In the inset, projection noise error bars are depicted. Using the same set of parameters, we calculate and plot the statevector simulation (from top to bottom: cyan stars with point up, diamonds and triangles with point up) for comparison. Another 20 $\times$ 100 shots on the quantum hardware are used for the classical SPSA optimizer to provide a search to update parameters for the next iteration step. Dashed lines show the ground state energy as determined by diagonalization of the Hamiltonian (gray), the average of the statevector calculation (blue) and the average of the trapped-ion processor (red) for the iterations where convergence is assessed to have been reached.}
    \label{fig:run-results}
\end{figure}

\begin{table*}[htbp]
    \centering
    \caption{Resulting energy expectation values}
    \label{tab:values}
    \begin{tabular}{|c|c|c|c|}
        \hline
        \textbf{Metric} & $\boldsymbol{K=-14}$ & $\boldsymbol{K=0}$ & $\boldsymbol{K=10}$ \\
        \hline\hline
        $\braket{W}$ exact & $-223.0$ & $-30.7$ & $1.0$ \\
        \hline
        VQE minimum, measured & $-215.8(3.6)$ & $-26.6(0.9)$ & $2.5(2.8)$ \\
        \hline
        VQE minimum, simulated & $-222.9$ & $-30.5$ & $1.3$ \\
        \hline
        VQE convergence distance $\Delta W$, measured & $22.6$ & $5.9$ & $11.3$ \\
        \hline
        VQE converged values spread $\delta W$, measured & $7.7$ & $1.2$ & $5.3$ \\
        \hline
        VQE convergence distance $\Delta W_\textnormal{sim}$, simulated & $2.1$ & $0.5$ & $1.3$ \\
        \hline
        VQE converged values spread $\delta W_\textnormal{sim}$, simulated & $1.9$ & $0.3$ & $0.8$ \\
        \hline
        Fidelity of reconstructed density matrix $\mathcal{F}$ & $0.66(2)$ & $0.70(1)$ & $0.61(2)$ \\
        \hline
    \end{tabular}
\end{table*}

The evolution of the cost function throughout the VQE iterations is shown for the different values of $\param = \{-14,0,10\}$ in Fig.~\ref{fig:run-results}. The energy values obtained on the quantum backend are shown along with energy values computed on a simulation backend for the circuit parameters used within each iteration. In all cases, convergence is observed within about 50 iterations. In the following, we discuss the properties of the VQE solutions within the converged regimes, where we observe energy  fluctuations between different iterations of $\delta W$ and mean offsets to the exact values of $\Delta W$. All values are listed in Table~\ref{tab:values}. The retrieved energy values match the exact values to within less than 10\,\% of the energy scale covered by the three phases. Thus, we find that while the measured energies display significant deviations from actual ground state energies obtained via exact diagonalization, the retrieved circuit parameters display a much better performance. When used on a simulation backend, the energies retrieved from statevector simulations using the optimum VQE parameters display offset $\Delta W_{\textnormal{sim}}$ reduced by about one order of magnitude. \\

For all three phases, only a small fraction of the observed fluctuations within the VQE iterations $\delta W$ is explained by residual changes of the circuit parameters, expressed by fluctuations of the exact values $\delta W_{\textnormal{sim}}$. Also, only a small fraction of the offsets $\Delta W$ are explained by imperfect VQE convergence, characterized by the residual offsets $\Delta W_{\textnormal{sim}}$, which are on the order of merely about 1\,\% of the overall energy scale. We find that the excess fluctuations $\delta W$ are statistically consistent with the shot noise errors  of the measured values as estimated via parametric bootstrapping (cf. inset of Fig. \ref{fig:run-results}). The offsets $\Delta W$ are on the one hand caused by shot noise, as sampling from a state close to a ground state can only lead to positive deviations of the retrieved energy. Also, errors on the two-qubit gates, leading to effective depolarization, contribute to the energy offsets. We also compute the fidelities of the tomographically reconstructed states with respect to the exact statevectors, for which values in the range between 60\,\% and 70\,\% are found~\cite{RawData}. 

Note that for obtaining the measurement data shown in Fig.~\ref{fig:run-results}, we have not applied any post-selection or other error mitigation technique, except for discarding invalid measurement results caused by loss of trapped ion qubits. 

For the three runs with different $\param$, the initial values of the parameters were chosen uniformly at random. However, the choice of their values strongly affect the convergence of the VQE algorithm. Moreover, one might observe some systematic effects of the different VQE runs, e.\,g. for the $\param = -14$ run (bottom), 160 iterations were recorded and the VQE shows a clear convergence with fluctuations becoming smaller over time due to the decreasing learning rate. For the VQE run with $\param = 10$, we observe initial fluctuations, which might originate from an intermediate failure of the control hardware. It can be seen that the algorithm resumes regular operation. This demonstrates that the VQE algorithm can compensate hardware failures to some extent. More interestingly, we observe a plateau of almost constant energy, maintained for about 20 iterations. This demonstrates the difficulty of finding a gradient for the classical optimizer. Moreover, this illustrates the difficulty of establishing meaningful convergence criteria for VQE. Here, we just run the VQE for a sufficiently long time and cross-validate with exact results.

\begin{figure}
    \centering
    \includegraphics[width=\linewidth]{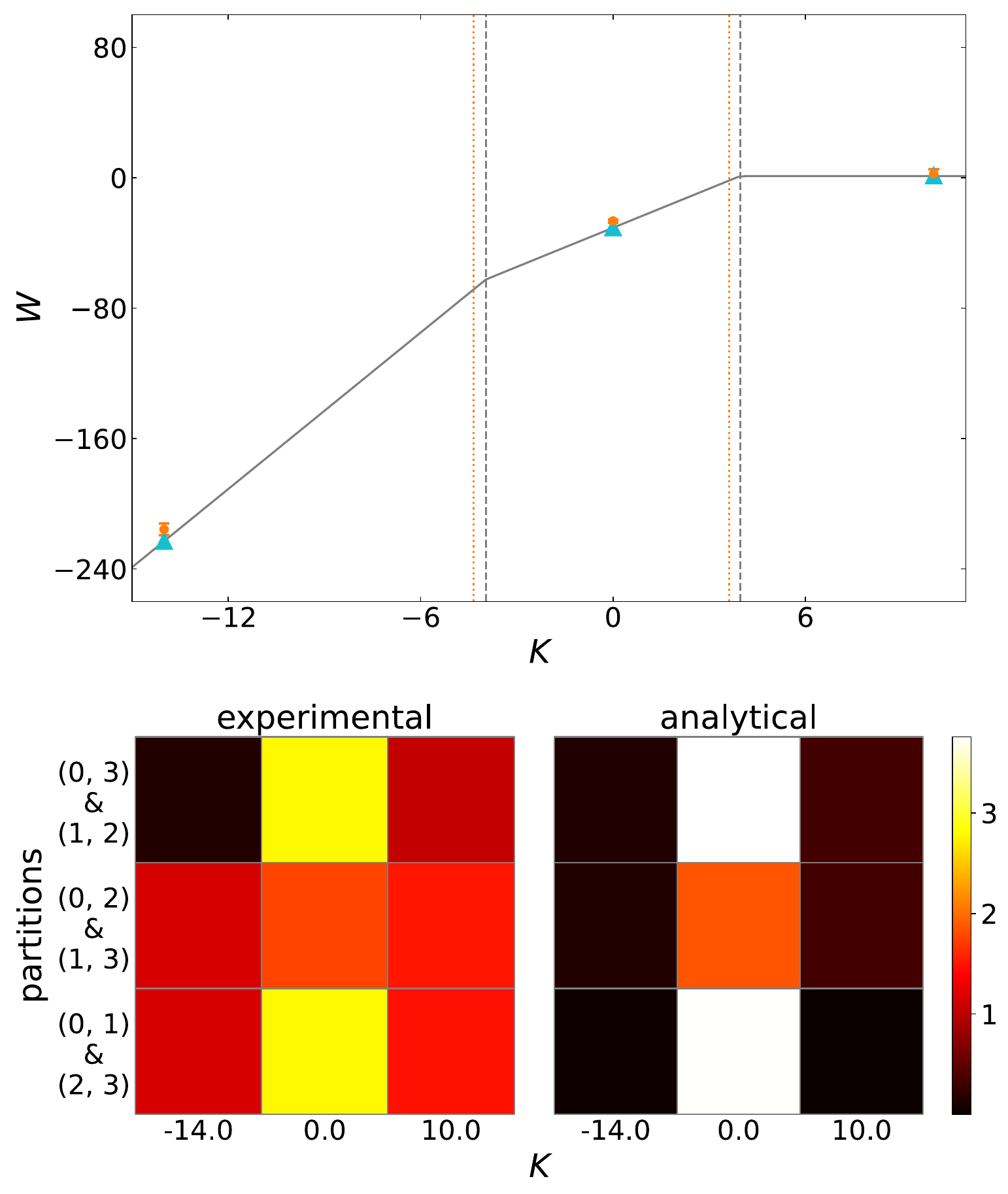}
    \caption[]{\textbf{Top:} Phase diagram showing the ground state energy as function of the chemical potential difference $\param = \kg$, scaled to the fermionic coupling $g$. The exact energy (gray line) exhibits a different slope within each phase. The minimum energy values $W$ from the trapped ion quantum processor results (orange circles, with error bars) are compared to exact calculation for the same parameters (cyan triangles). The phase boundaries inferred from the measured (orange dotted line) and the exact calculation (gray dashed line) are shown as well. \textbf{Bottom:} Quantum mutual information between different two-by-two partitions for the exact (right) and reconstructed density matrices (left). The heatmaps show the values for different qubit partitions (vertical axis) and the different values for $\param$ (horizontal axis).}
    \label{fig:phase-diagram}
\end{figure}

\begin{figure*}[htbp]
    \centering
    \includegraphics[width=0.8\linewidth]{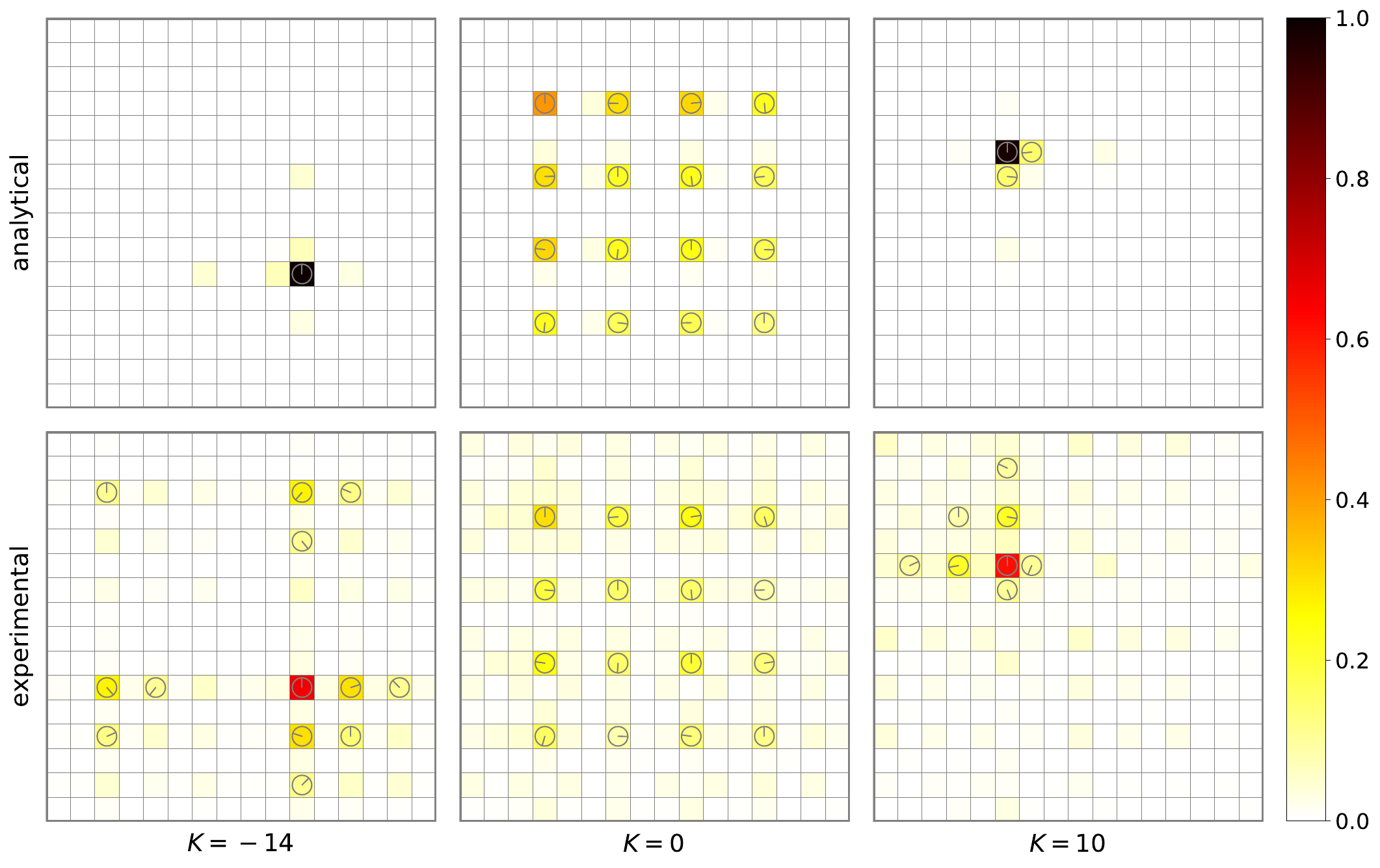}
    \caption{Reconstructed density matrices for tomography results for different values of the system parameter \param. Depicted are the absolute values of the density matrix with indicators of the phase for significant entries. From left to right and top to bottom, the entries are ordered from computational state $\ket{0000}$ to state $\ket{1111}$ with $q_0$ as the most significant bit. Each basis of the state tomography for the experimentally obtained density matrices was measured with 400 shots.}
    \label{fig:tomos}
\end{figure*}

For a final assessment of all three VQE runs, we take the lowest energy $W$, as evaluated by the quantum backend for each run, and plot this together with the corresponding exact energy value against the system parameter \param, see Fig.~\ref{fig:phase-diagram} top. To calculate the phase transitions from the VQE result, the mean particle number $N_f$ for a specific flavor $f$ is determined by an additional measurement of the corresponding circuit with the observable \paulistring{ZZZZ} and application of Eq.~\eqref{eq:particle-number-by-flavor}, leading to $N_f=\{1.73(2), 0.97(3), 0.17(2)\}$. To obtain physically meaningful values complying with the charge constraint $Q_{\text{tot}}=0$, we round the values to integers and compute the phase boundaries using  Eq.~\eqref{eq:MFS_crit_pt}. The quantum backend results of $-4.3(5)$ and and $3.6(4)$ and the exact results $-3.96$ and $3.96$ are indicated in Fig.~\ref{fig:phase-diagram}. The experimentally obtained values for the phase boundaries agree with the exact values within one standard deviation.

In order to analyze the properties of the approximate ground state retrieved by VQE, we quantify correlations via the computing the quantum mutual information (QMI) $I_{A:B}(\rho)=S_A(\rho)+S_B(\rho)-S_{AB}(\rho)$ from the reconstructed density matrices (cf. Fig.~\ref{fig:tomos}) $\rho$, where $A,B$ is any of the three possible two-by-two bipartitions of the four qubits and $S(\rho)$ is the von-Neumann entropy evaluated for either a subsystem or the entire register. We show a comparison of the QMI values computed from exact and tomographically reconstructed density matrices for all three phases and all three bipartitions in Fig.~\ref{fig:phase-diagram}. For $K=0$, we see large QMI values indicating significant correlations between the bi-partitions $(0, 1)/(2, 3)$ and $(0, 3)/(1, 2)$, respectively. Mapping this back to the staggered lattice, this translates to high correlations between particles present at the lattice sites $n=0$ and $n=1$ (cf. Eq.~\eqref{eq:JWqubitmapping}) related to particle / anti-particle excitations of the same flavor on those lattice sites. Additionally, we observe reduced but nonzero correlations in for bipartition $(0,2)/(1,3)$, which translates to correlated creation of particle pair of different flavor $f=0$ and $f=1$. These correlations result from the interaction terms of our model Hamiltonian $W$, which are dominating in this phase since $K=0$ translates to $\nu_0=0$ as $\nu_1=0$ was initially fixed (cf. Eq.~\eqref{eq:hamiltonian_dimensionless}). For $K=-14,\, +10$, we see a vanishing QMI in all bipartitions for the exact density matrices. This indicates a lack of correlations between any bipartition due to the dominating second term of $W$ in these two phases (cf. Eq.~\eqref{eq:hamiltonian_dimensionless} with $\lvert\nu_1\lvert = 112, 80 \gg x=16$). This term is local and consists of single-qubit $Z$ operators without any interaction.

\section{Conclusion and Outlook}
\label{sec:assessment}

In this work, we have demonstrated a fully automated execution of a VQE algorithm on a trapped-ion based quantum computer backend for characterizing phase transitions in the two-flavor Schwinger model with chemical potential employing staggered fermions. We emphasize that a full VQE as a hybrid setup comprised of a classical optimization algorithm and a quantum backend has been executed. The motivation for this work has been that the physical situation of adding chemical potentials is very difficult to be realized with classical Monte Carlo methods, since it leads to a sign problem in the action formulation. In our work, we could show that formulating the problem as a VQE task circumvents the sign problem and that indeed the phase transitions of the model can be mapped out. 

Our results also allow serve as a benchmark for trapped-ion based backends in the context of VQE: The possibility to maintain relatively high operational fidelities over the course of entire VQE runs allows for determining the rather accurate circuit-parameters, while the comparatively long data acquisition times render the retrieved absolute energy values to be prone to shot noise. With seven free circuit parameters to be determined by the VQE algorithm, with a total run time of about three days, or about 750\,000 shots in total, our experiment proves the long-term stability of the setup and validate the automated  calibration routines for a concrete use-case. Future challenges of the trapped-ion hardware will be the scaling to higher qubit numbers and improved gate operations, involving more advanced error suppression techniques, and a reduction of the effective duration per shot by optimizing software latencies and reducing operational overheads.

Our experiments demonstrate the viability of using VQE on a quantum backend to address problems in condensed matter and high energy physics that defy employing classical Monte Carlo methods. This approach also opens new paths to investigate other directions which cannot be tackled by classical methods: This includes topological terms which have been worked out in 2+1-dimensional~\cite{Peng:2024xbl} and 3+1 dimensional~\cite{Kan:2021nyu} quantum electrodynamics already and which would allow to develop new materials or to study  CP-violation in the context of high energy physics. In addition, chemical potentials play an important role also in higher-dimensional theories. Another promising direction are quantum simulations of real-time dynamics, e.\,g. scattering experiments or the dynamics concerning phase transitions and quenches.

Acknowledgments: The JGU team acknowledges financial support by the BMBF within the projects HFAK, IQuAn and ATIQ.  We thank Joachim v. Zanthier for helpful discussions at an early stage of the project and for establishing this collaboration. This work is funded by the European Union’s Horizon Europe Framework Programme (HORIZON) under the ERA Chair scheme with grant agreement no.\ 101087126 and by the Ministry of Science, Research and Culture of the State of Brandenburg within the Centre for Quantum Technologies and Applications (CQTA). 
\begin{center}
    \includegraphics[width = 0.08\textwidth]{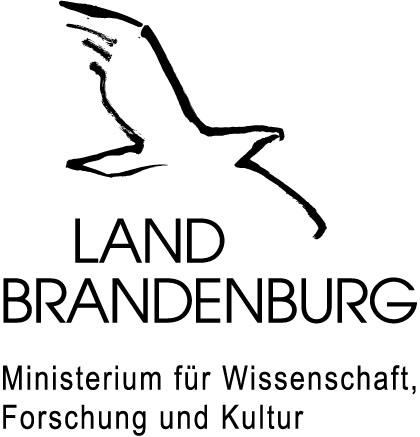}
\end{center}

\bibliography{papers}

\begin{thebibliography}{45}
\bibitem [{foo()}]{footnote1}%
  \BibitemOpen
  \href@noop {} {}\bibinfo {note} {The resulting SPSA hyperparameter used are: i) perturbation sequence of $\frac{0.2}{(1 + k)^{0.101}}$ and ii) learning rate sequence of approximately $\frac{0.03174}{(1 + k)^{0.602}}$ for iteration index $k$.}\BibitemShut {Stop}%
\bibitem [{Raw()}]{RawData}%
  \BibitemOpen
  \href@noop {} {}\bibinfo {note} {Data is available from the authors upon request}\BibitemShut {NoStop}%
\end{thebibliography}

\end{document}